# SOCIAL MEDIA HARM ABATEMENT: MECHANISMS FOR TRANSPARENT PUBLIC HEALTH ASSESSMENT

Nathaniel Lubin[1], Yuning Liu[2], Amanda Yarnell[3], S. Bryn Austin[4], Zachary J. Ward[5], Ravi Iyer[6], Jonathan Stray[7], Matthew Lawrence[8], Alissa Cooper[9], Peter Chapman[10]

## ABSTRACT

Social media platforms have been accused of causing a range of harms, resulting in dozens of lawsuits across jurisdictions. These lawsuits are situated within the context of a long history of American product safety litigation, suggesting opportunities for remediation outside of financial compensation. Anticipating that at least some of these cases may be successful and/or lead to settlements, this article outlines an implementable mechanism for an abatement and/or settlement plan capable of mitigating abuse. The paper describes the requirements of such a mechanism, implications for privacy and oversight, and tradeoffs that such a procedure would entail. The mechanism is framed to operate at the intersection of legal procedure, standards for transparent public health assessment, and the practical requirements of modern technology products.

1 The Berkman Klein Center, Harvard University

2 Center for Health Communication, Harvard University TH Chan School of Public Health

3 Center for Health Communication, Harvard University TH Chan School of Public Health

4 Harvard T.H. Chan School of Public Health

5 Harvard T.H. Chan School of Public Health

6 Neely Center, University of Southern California Marshall School of Business

7 Center for Human-Compatible AI, UC Berkeley

8 Emory University School of Law

9 Knight-Georgetown Institute, Georgetown University

10 Knight-Georgetown Institute, Georgetown University

## INTRODUCTION

In recent years, news headlines have featured allegations of the harmful effects of social media – including international genocides[1], promotion of health conspiracy theories during the COVID pandemic[2], and threats to democracy[3,4]. For all the coverage and apologies from company executives[5], one set of harms has consistently generated broad support for the need for mitigations: harmful effects on the wellbeing of children[6]. Those harms include mental health effects stemming from social comparisons[7], reduced educational attainment[8], excessive usage that displaces activities like sleep[9], and more acute issues like harassment and sex trafficking[10].

Governments have begun to act in response to these alleged harms, resulting in a number of new state laws and a bevy of proposed federal action[11]. Yet, partisan gridlock and legal hurdles suggest that legislative activity may take time to materially reduce harms[12]. There is significant regulatory activity targeting social-media-related harms in countries around the world, including the United Kingdom[13,14], European Union[15] and Australia[16]. Even as regulation and broader societal concerns are increasing, many companies have scaled back the size and scope of trust and safety teams[17] and restricted external oversight[18].

As the prospect of federal legislation in the US remains uncertain, litigation has pushed forward. Attorneys general in more than 40 states from across the American political spectrum have brought lawsuits against social media companies[19,20]. Federal lawsuits are pending over children's data, privacy, and online safety[21], including a currently-pending consolidated multidistrict litigation in California[22] and its counterpart coordinated civil cases in state court in California[23]. School districts[24], advertisers[25], tribal nations[26], and news providers[27(p. 200)] all have filed suits against social media companies.

This reality mirrors the experience of other industries alleged to have caused significant harms, notably including tobacco and opioids. In those cases, product liability claims led not only to large financial payouts[28], but also to significant changes in marketing rules[29], access and age restrictions[30,31], and much better documentation of adverse health effects[32,33].

While social media cases are still pending, we believe there is a need for a public discussion of strategies for remediation to ensure that, in the event that companies are required to come into compliance via a court order or settlement, concrete steps are taken to mitigate the harms of products. It may appear premature to develop a plan prior to the conclusion of a specific case. Yet a remediation regime capable of meeting the complex needs of such a situation would benefit from advance thinking given the likely short timeline a judge would be on while either reviewing a settlement agreement or trying to scope a judgment in the short period following a finding of fact.

Anticipating this challenge, in this article we offer a proposal for how an abatement plan might be established. Our mission is to develop a mechanism that (1) is legally enforceable, with reasonable precedent from other industries; (2) has an established evidence base in the public health literature; and (3) is implementable technologically based on the product systems and procedures that social media companies operate within. Of course, any particular case will incorporate myriad specific fact patterns and details including information garnered during discovery, requiring updates and further specifications including direct rules or obligations to limit specific harms. We do not argue for a blind implementation of any particular structure but rather show that this approach is possible and, as precisely as we can, offer specifics for how this could be done.

Below, we first situate this framework within legal, public health, and technology contexts. We then offer a detailed description of the proposed mechanism itself, including its goals, implementation requirements, benchmarks, and specifications for oversight and management. Next, we discuss the metrics suitable for supervision. We then offer a brief discussion of some of the key questions posed by this framework, including potential risks and drawbacks and associated ethics questions, before concluding.

## CONTEXT

Social media is, of course, not the first product or service to raise public health or consumer protection concerns. Nor is it the first to see the courts looked to as a potentially-central player in crafting a regulatory response to such concerns. So while public documentation for implementing an abatement tied to social media does not yet exist, there is a deep history in the legal, public health, and technology communities to inform how this process may be pursued.

### Legal

Public health law is replete with prior examples of intervention on behalf of health concerns. The multi-state tobacco settlement negotiated between 46 states and cigarette manufacturers remains an important source of regulation on cigarette manufacturing and sales[34]. Judge Gladys Kessler's subsequent order in the RICO case brought by the DOJ, *United States v. Philip Morris USA* in 2006, is another[35]. A court-overseen settlement between states and Juul in 2023 has played a primary role in regulating e-cigarette use[36]. A settlement between states and opioid distributors regulates opioid sales[37], as do related judgments against opioid manufacturers[38]. While the scope of specific remedies may differ within a negotiated settlement and a court order, both avenues likely offer near term paths for court-mandated limits on social media.

These precedents illustrate reasons that court-based remedies—through both court order or negotiated settlement—can be a useful public health tool as a supplement or alternative to law and regulation imposed by state or federal regulators[39]. Court-based approaches allow for regulatory protections to be negotiated between industry and representatives of persons most directly harmed by past industry actions, two core constituencies. In a negotiated settlement, this approach also allows for the imposition of protections that would otherwise be contested as unconstitutional or impossible. For example, the tobacco master settlement included marketing restrictions similar to those that had previously been invalidated by courts when imposed by state law[40]. The tobacco companies' agreement to the court-based remedy circumvented litigation about constitutional limits on regulation that could have prevented a state legislature from imposing the same remedy outside of the litigation context. These considerations are particularly salient in social media where multiple courts have preliminarily enjoined some state social media regulations, including those mandating broad risk assessment of privacy risks for children[41], mandating reporting on content moderation[42], and some special design restrictions on minors' accounts[43].

The role of court-based remedies in addressing prior public health concerns posed by industry makes it reasonable to predict that a court order or negotiated settlement structure will play a central role in shaping the future of public health limits on social media. The technology industry already has a long tradition of court-negotiated settlements and consent decrees with the Federal Trade Commission (FTC), supported by state attorneys general, that have required product and governance changes related to privacy, security, and harms to children[44–51]. The fact that courts have rejected social media companies' motions to dismiss in multiple pending lawsuits—allowing cases into discovery and then merits briefing—also provides reason to believe that court-based mediation of social media harms is likely[52].

At the same time, past experience with court-based public health remedies reveals dangers. Studies of the tobacco master settlement raised serious concerns, documented in the literature, that the agreement did not do enough to actually curb smoking-related harms[53]. More recently, provisions of the opioid distributor settlement have been blamed, paradoxically, for impeding the distribution of medications for the treatment of opioid use disorder[54]. One obstacle to successful court-based remedies is the fact that the development of such rules are not subject to the same open, participatory, expert-inclusive processes as lawmaking. Court-based remediation understandably centers the parties to a particular case, yielding fewer opportunities for evidence-based policymaking and vetting of potential regulatory provisions by the public and experts.

In short, court-based remedies can play a major role in the future of social media, but the constraints of judicial interventions may impede the incorporation of public and expert input into negotiations about such remedies. For these reasons, there is significant value in scholarly consideration now, *before* cases have reached later stages and outside the context of any particular negotiation or proceeding, of what court-based remedies addressing social media harms might entail. Just as peer-reviewed research proposing and evaluating state and federal legal requirements helps lawmakers develop and implement evidence-based policy in the course of lawmaking, we believe that peer-reviewed research proposing and evaluating a court-based regime for limiting social media harms would help judges and the parties to court cases in developing and implementing evidence-based reforms in a judicial setting.

## Public Health

Public health experts have traditionally relied on a variety of empirical study designs to evaluate the health risks of real-world use of products like tobacco. For example, longitudinal studies, in which scientists observe tens of thousands of individuals' smoking habits and health outcomes over many years, have linked cigarette smoking to increased mortality rates from conditions like lung cancer, coronary artery disease, and chronic obstructive pulmonary disease[55,56]. Case-control studies, in which scientists compare the health behaviors of cancer patients (cases) with healthy individuals (controls) over time, have shown a dose-response relationship between greater smoking intensity and duration and increased lung cancer risk[57]. And biomarker studies, in which scientists precisely quantify the amount of nicotine metabolites in the lung tissue of smokers and nonsmokers, have allowed precise measurements of exposure from tobacco smoke[58]. Together these kinds of observational studies have established clear causal links between tobacco use and various health harms.

But these kinds of observational studies are challenging to apply to social media for at least two reasons. First, scientists lack access to the types of data suitable for rigorous understanding of how platforms' personalized and ever-optimized algorithms shape individuals' social media experiences. Without such data, one user's social media experience cannot readily be compared to another. Second, the effects of interest are often long term, unevenly distributed, and operate at the level of ecosystems. The algorithms used for content selection both influence and are influenced by users, users strongly influence each-other, and there are complex interplays between the demand, supply, and distribution of various types of potentially harmful content.

The gold standard in assessing causality in public health is the randomized controlled trial (RCT). RCTs complement a range of relevant methodologies for assessing short and medium term impacts[59]. In an RCT, participants are randomly assigned to either a treatment or a control group, allowing scientists to detect stronger evidence of causation than in observational studies. In the case of tobacco and opioid harms, RCTs have been used to track the effectiveness of smoking cessation and harm reduction interventions[60].

Public health agencies also have a long history of studying the incidence of harmful behaviors over time. The CDC regularly runs the Youth Risk Behavior Survey[61] which examines reports of sexual behavior, substance use, and mental health. Such surveys have begun to include social media usage, but those measures have been sporadic. In the case of social media, there remains a need for similarly rigorous, longitudinal evaluation of harms[62]. Independent researchers have recently conducted RCTs to track the effectiveness of various harm reduction interventions, including limiting time spent on social media[63,64], coaching users on how to minimize negatives and maximize the benefits of social media use[65], and increasing the amount of evidence-based content on social media platforms[66]. And platforms run RCTs on their products regularly, and some results of experimental studies by platforms have been used to inform potential mechanisms for reducing harm[67,68].

But social media companies provide very limited data access related to the workings of their platforms and algorithms, and make on-platform RCTs very difficult for independent scientists to access or conduct, except for limited active collaboration with platforms[69]. The existing tools that researchers can access due to these restrictions are ill-suited to causally assessing long-term effects of inherently transient digital interactions at scale or untangling how potential harms might affect specific user subpopulations differently, especially over longer time horizons. The lack of access to platforms' internal experimental testing and longitudinal data from users hinders the ability of external scientists, investigative journalists, watchdog groups, and regulators to fully examine plausible causal links between social media and harms. In response, some jurisdictions have begun mandating access to such experimental testing[70,71].

Courts instituting previous abatement remedies have required that the makers of products found to cause harm enhance data access and quality. It would thus allow for courts to require data access, including experiments, that could be used to monitor for the reduction of specific social media harms on those most at risk. As detailed below, this would also include tracking of intermediate indicators related to user experience, psychological/neurocognitive responses, and behavioral risk factors that are predictive of longer-term health outcomes.

## Technology

The dominant paradigm for technology companies tracking their negative effects relies on definitions of specific categories of harm. Companies have policies, sometimes referred to as community standards or guidelines[72], that define content and behavior that is not allowed on the platform. These definitions enable measurement of how prevalent content and behavior that fits those definitions is. By randomly sampling a set of content or behavior on the platform, one can enable trained raters to determine how often people view content that violates a given policy, or how many times accounts are taken down for a given behavior. Teams focused on trust, safety, and/or integrity at companies work to develop new policies to fit emerging harms, develop new processes and products to reduce the prevalence of defined harms, and generally measure their progress against the prevalence of violating content and behavior. Internally, these metrics are used in randomized control trials to understand whether new products and processes do or do not reduce harmful content and behaviors. Externally, these metrics can be used to report progress to users, regulators, and other societal stakeholders[73].

Typically, the prevalence estimates reported by companies do not match the magnitude of negative experiences that people in society report. For example, in one report, Facebook reported that only 0.08% of views of content violated their bullying and harassment policy, yet surveys indicated that 11% of 13-to-15-year olds report being bullied within a seven-day period[74]. Civil society groups often report violating content to platforms and receive official responses that such content does not violate policies[75]. Court filings further suggest that platforms often do not take enforcement actions against reported content that many in society consider harmful[76].

Because stakeholders outside of companies lack the ability to randomly sample content and behavior on the platform, they may create test accounts (usually in violation of terms of service[77,78]) and report on those experiences[79–81]. They may also take reports from members of their community who report negative experiences[67] or they may observe viral content that is harmful[82]. These methods often suggest that harm is more prevalent than platform metrics, which are generally based on platform policy definitions of what is or is not explicitly violating, would suggest[83]. This disconnect—between external observations of high levels of harm and policy based platform metrics that show relatively low levels—has been highlighted by former technology workers[84] as well as civil society groups[85], who note that policies may not adequately capture experiences of harm. Technology companies have acknowledged that "borderline content" that does not fit existing policy definitions of harm may often get more engagement and therefore be promoted by the platform's engagement-based recommender systems[86]. Such engagement-based optimization makes such borderline content appear more often in users' feeds.

To fill this gap, most companies also attempt to understand "bad experiences." In one leaked Facebook study, more than 70% of users reported seeing content they wanted to see less of, regardless of whether such content violated company policies, and they reported often seeing it within the first 5 minutes of scrolling and multiple times per session. As part of discovery in one lawsuit[87], it was disclosed that Instagram conducted regular assessments of users' negative experiences. In surveying a random sample of users as to their bad experiences, the company found that, in a seven-day period, 19% of 13-15 year olds reported getting unwanted sexually explicit content, 21% reported seeing content that made them feel worse about themselves, 13% received an unwanted sexual advance, 11% reported being bullied, 8% reported seeing content that encouraged or depicted self-harm, and 4% were offered drugs. Given the number of teens on Instagram, these numbers suggest that millions of teenagers are having acute persistent negative experiences on Instagram.

Unlike platform prevalence metrics that require the ability to randomly sample content or behavior, user experience surveys can be conducted both internally and externally to platforms. Several regulators and civil society groups have tracked negative experiences within their populations. For example, USC's Neely Center has conducted surveys using questions that were previously used at companies and shown that adding a platform-specific dimension can allow for cross-platform comparisons that match other data sets. Those surveys have shown how X/Twitter users often are exposed to harmful content, matching studies conducted by nonprofit organizations[88]. They have also shown that the platform Nextdoor creates a disproportionate number of negative experiences relative to other digital sources, matching both internal and external analyses[89], and providing data points for internal advocates to leverage in advocating for product changes. Work is now ongoing across jurisdictions, such as the UK's online experience tracker[90], to develop surveys that can help with longitudinal, cross-platform studies of negative experiences.

For use as a product team goal within companies, user experience metrics have limitations in that they are harder to collect than measures of direct engagement, and their subjectivity makes them more difficult to calibrate. Metrics based on content and behavior are likely to be more sensitive to interventions. Still, they provide an important external validation to metrics that anchor on companies' differing definitions of what is or is not allowed on their platforms. Most important, for external parties, they provide a feasible method to understand platform specific harm over time, including in relation to court mandated product changes. To that end, and considering lessons from tobacco, opioids, and other prior cases, we develop below a proposed framework for court-based regulation of social media.

## MECHANISM

A suitable abatement mechanism has several requirements. It must legally operate within the scope of a decision or settlement, representing an appropriate response to mitigate the scope of a finding of fact. Yet, it must be ambitious enough to substantively deliver actionable data within the timeline of remediation, likely many years, following a decision[91]. More tangibly, there must be specific potential harms to measure and report on. Achieving such an end requires transparency for the public, but also privacy for individuals whose data contribute to ongoing assessment.

In what follows, we assume the mechanism is in response to a case associated with the health effects on children, such as the pending federal multidistrict litigation[92] cases against social media platforms[28]. These cases allege personal injury to members of the public stemming from negligence and product liability caused by usage of platforms' suite of applications. Allegations include harms associated with addiction and lack of sleep, harmful comparisons to other users, facilitation of harassment, and the furtherance of other crimes including the distribution of CSAM and sex trafficking. We focus here on some of those allegations more than others, both because we cannot fully describe harms prior to specific case selection, but also because we deliberately choose a framework suitable for some of the likeliest concerns should these cases prove successful. This mechanism is designed to operate *in addition* to whatever specific requirements the court may order so as to ensure compliance and efficacy by the platforms in their efforts at remediation.

A review of court cases and legislation[93] addressing online harm suggests that harms can be grouped into four broad categories of unwanted/harmful content, unwanted and excessive usage of platforms, unwanted/harmful contact, and unwanted/harmful usage of a users' information and likeness (see Appendices for further details). While we cannot cover all harms that may be covered in all cases, we cover each of these categories in our proposed mechanism and would assume that similarly situated harms from other cases would likely benefit from similar mechanisms.

Though there is significant public consensus on the need for intervention to protect children[94] from real and demonstrable harms, there is heterogeneity in the academic literature over the size and scope of effects[95]. Nonetheless, a well-designed protocol should yield further and longer-term causal evidence on how the specific design practices of platforms lead to specific experienced harms. It should also produce evidence on more general, long-term outcomes such as well-being or depression. Because a great many off-platform life experiences influence these higher-level outcomes, it is more difficult to measure platform effects and they are less likely to change with platform design changes. Nonetheless, direct measurements of well-being and mental health using survey instruments are of great interest, and we also recommend assessing how these measures relate to more immediate user experiences of harm. In the course of this measurement, the plan sets requirements for greater transparency and access to information about the steps companies take to mitigate harms and how they set goals.

**Operating structure**

Tactically, a system capable of assessing harms on at-risk users over a period of years must allow for:

1. Measurement of harm for a representative population over a period of years;
2. Adequate assessments of product or operational changes by the company during the period of observation, including how changes experimentally affect measures of harm;
3. Data and benchmarks to allow for normalization to account for changes in the world external to the company/product;
4. Assessments of subpopulations where effects are alleged to be largest (such as among particularly high-engagement users), preventing overall averages from hiding unacceptable effects.

Given these requirements, we propose a two-pronged structure, with one part focusing on internal company data and actions that only the company can execute (what we call the "internal mechanism") and a second part (the "external mechanism") focusing on external experiments and analyses that do not depend on company access or compliance.

The internal mechanism can access and employ true (randomized) experiments to estimate the causal impact of design changes on outcomes that are more sensitive to changes over a short time period (e.g., user behavior and user experiences). We also suggest that the company maintain a recurrent "holdout" group of users who do not receive updates to the platform design for some extended period. This will serve as a crucial control group for a variety of analyses and assessments. An internal assessment can also leverage access to compare outcomes with the goals, "objectives and key results" (OKRs), and benchmarks of different teams across the company[96].

The external mechanism would collect a stream of observational data by surveying platform users. This external assessment can further enable cross-platform comparisons, and contribute to scientific knowledge on the relationships between social media use, intermediate outcomes, and indicators such as subjective well-being that evolve over a longer timeframe (e.g. 6+ months) than is typically observed in on-platform experiments (typically weeks at most)[97,98].This proposed two-pronged structure thus leverages both experimental and observational designs to assess a continuum of outcomes in a transparent and robust approach.

**Figure 1:** Usage of the internal and external mechanisms

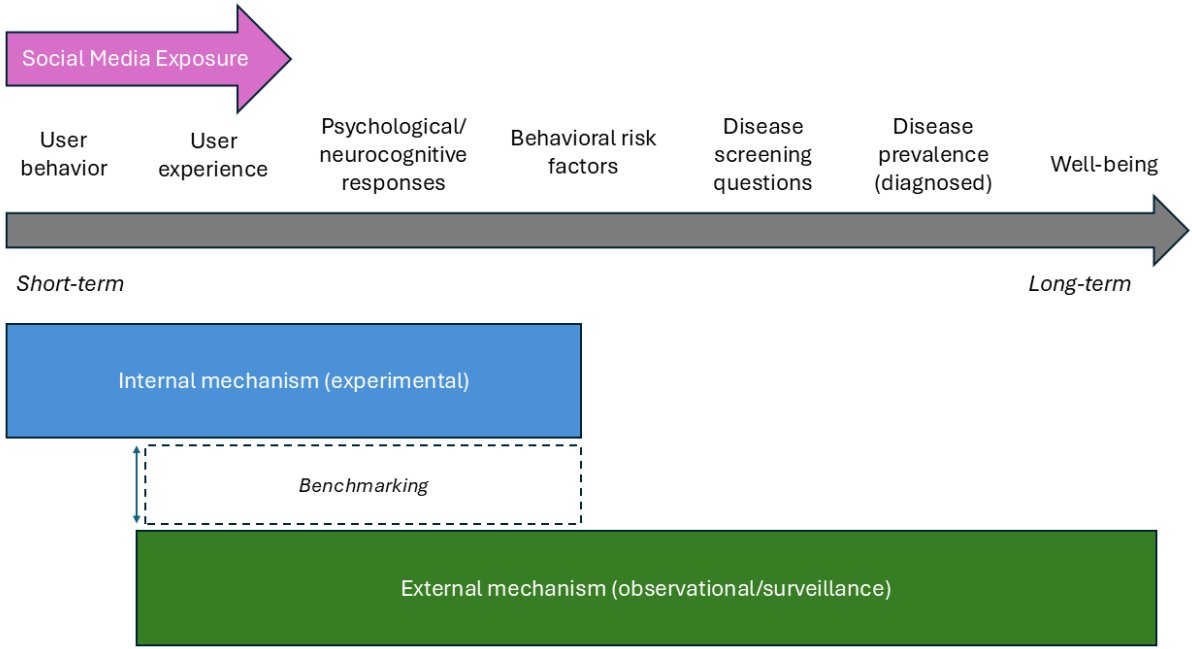

For both the internal and external mechanisms, groups of particularly high-risk users will need to be identified and oversampled. In general, specific subgroup (e.g. neurodivergent youth[99]) requirements must be determined based on the details of a case. Yet there are few subgroups we can predict will be necessary in almost any potential case. One group for oversampled evaluation will be usage-based: metrics will need to be reported and assessed when segmented by time-based engagement, with extra emphasis on the most-engaged cohorts of users (e.g. the top 1% or 0.1% of users). Additionally, oversamples are likely to be necessary for groups of users who are most at risk from exposure based on the literature, including users of color and LGBTQ users[100–102]. Such analysis invariably will raise privacy risks and the abatement plan will clearly spell out necessary mechanisms to ensure privacy is protected.

**Internal mechanism**

The internal data system focuses on the product and operational decisions made by the company. It is designed to lean into the advantages afforded by direct coordination with product development as part of the legal process. For the duration of supervision, the internal system will have three requirements, with results reported back to the court intermediated by an external group of experts (discussed more below). A throughline of these components is that assessments and requirements follow the existing product evaluation and experimental procedures that companies *already* use[103]. A judge implementing this plan would be wise to make systems/plans robust enough to take advantage of already-created optimization systems while ensuring sufficient external review and supervision.

The three components of the internal system include:

- Universal holdout: Large technology companies such as social media platforms make thousands of parallel product changes across numerous engineering and product teams. While companies usually A/B test changes, it is often difficult to understand aggregate effects over time precisely because of the complexity of these systems and the interaction effects between updates[104]. A cornerstone of the internal system is that companies will need to create universal "holdout" groups that receive no new product updates for some duration, so as to assess whether the aggregate decisions made by the companies are actually improving outcomes – or at the very least not making products worse. (Changes which would be unethical to withhold can be exempted from the holdout.) Comparisons to the holdout group should be designed to mirror the engineering timelines and internal product evaluation cycles, which are typically six or twelve months.

- Operational and product experiment transparency: To garner insight into the actual decisions made by the operational leaders in the company, a second component of the internal system is to provide external access to the goals and product experimentation data that product teams implement and are judged against. This would include insights into the product roadmap as well as an ongoing inventory of experimentation and the results of each experiment, including measures of harm[105,106]. Because the choices made in response to these experiments often involve evaluating tradeoffs (e.g. usage growth versus risks of abuse), understanding how and why products shift is central to ensuring abatement of abuse. The point is not that every experiment needs to be implemented to mitigate abuse but rather that the overall set of decisions should incorporate greater emphasis on potential harmful externalities. By offering transparency into these goals and results, even without further obligations, monitoring can introduce long term accountability by providing insight into how monetization is weighted versus other considerations. Besides the experimental data, this transparency includes the management structure around experiments and how product decisions are made, including product team and organization goals, reflective of decision making strategies that companies often utilize, such as metric goals and OKRs[96].

- Added Metrics: For experimental protocols, a critical requirement will be the inclusion of impact metrics designed to assess effects associated with harms to audiences identified by the relevant litigation. Although typical health metrics may not be sensitive to short-term changes in platform design, a set of intermediate indicators including user experience, psychological/neurocognitive responses, and behavioral risk factors that are predictive of longer-term health outcomes can be routinely assessed by experimental protocols. (More details about metrics are included in the next section and in the Appendices.)

The settlement could further mandate external transparency of internal studies and documentation related to relevant harms. The tobacco Master Settlement Agreement, for example, mandated disclosure of internal documentation surfaced during litigation[107].

## External mechanism

The external system enables data validation and comparisons through the collection of information on many outcomes over a longer time period. This system mirrors existing public health practices that monitor usage of tobacco or other risky behaviors[108]. While internal data are critical for understanding the decisions and operations made by actors within the company, it is inherently subject to internal logistical challenges and will struggle to provide transparency about the broader effects of a platform, including comparisons to other companies over time. By establishing a consistent data collection process for external assessment of user experiences and views, in parallel to internal assessments, the combined system will be able to benchmark shifts and better normalize and understand effect sizes. Additionally, because data will be sourced independently from preexisting company data, affording opportunities for expanded participant consent at the point of collection, an external system would not be subject in the same way to the legitimate user privacy issues of the internal system, making it easier to over-sample vulnerable subgroups and estimate more granular results. It would also be more independently auditable such that society would have more confidence in this data.

The external system would include the following components:

- Longitudinal user studies: The cornerstone of the external system will be an ongoing, periodic set of large-sampled public research with special focus on at-risk populations about their experiences online related to the relevant company as well as competitors. These surveys would operate similarly to the BEEF ("Bad Experiences and Encounters Framework")[87] survey that Meta at one point implemented to understand questions related to topics like harassment, negative associated affect, and exposure to inappropriate material[109]. Numerous other companies and jurisdictions have adopted similar forms of user experience measurement. Periodically, these users would also be assessed using longer term health metrics, that can help connect discrete user experiences to longer term outcomes of interest.[11]

- Cross-platform comparisons: One benefit of an external system is that each platform's results can be evaluated by using parallel products as benchmarks. An advantage from inside knowledge of the internal system is that the group overseeing external analysis would have direct knowledge of timelines for new products or feature rollouts from the target company (including specific populations receiving the changes). In practice, the scope of this analysis would include measures of user experience of parallel products, especially other social media and technology products. The fact that some platforms' users report far higher levels of harmful experiences as compared to users of other platforms, should motivate deeper examination of high-harm platforms.

---

[11] If participants in longitudinal studies have sufficient overlap with the internal platform holdout, and if sufficient privacy protections can be enacted, it may also be possible for this infrastructure to be directly benchmarked against the internal mechanism. Under the right circumstances, strong causal inference methods could then be used, such as difference-in-difference analysis comparing holdout and non-holdout users before and after a product change. Essentially, this would make it possible for external researchers to test the effects of recent platform changes on any outcome which can be observed without platform data (e.g. from survey data). Such a structure would *not* be necessary for the external mechanism to serve its core utility, but would strengthen the infrastructure.

- <u>Establishing access to samples of popular content</u>: With the rise of social media litigation in the US and regulation in Europe, platforms have taken steps to restrict access to public social media data even as regulatory requirements expand in Europe. For example Meta[18,110], Twitter/X[111], and Reddit[112] have all narrowed access to publicly available data. To the extent possible based on the material accessible, the external mechanism would facilitate researcher and public access to public social media data. This language would be informed by existing regulatory requirements in the EU[113] as well as proposed language in federal legislation[101,102] that seek to advance journalism and public interest research while ensuring user privacy and platform integrity. Affirming public access to data would serve as an important complement to off-platform assessment and advance two important abatement goals: more informed public debate and more effective external assessment. A specific starting point for public access should include real-time visibility of public content available on the platform, sortable by how much engagement each piece of content has gotten, including user reports, and by the presence/absence of particular keywords. This kind of data has already proven useful for holding platforms accountable both internally[116] and externally[18].

**Benchmarking and Oversight**

Implementing the system envisioned here requires active participation from the court or a court-ordered overseer such as a state attorney general, the company, and a group of external researchers[117,118]. The court must sufficiently empower these external researchers to establish and implement the external mechanism, and to direct, access, and review the goals/management documentation and experimental results (but not all the private data outside of auditing) for the internal mechanism. This type of work is likely to be significant and substantive, spanning several years, meaning it will be critical for the court to allocate sufficient funding for the program as part of any financial abatement/settlement.

Much like the group of authors collaborating on this article, the court-appointed external research group should consider expertise across a range of disciplines. In structuring and determining appropriate resourcing for the external research group, the court should leverage insights from previous court-appointed public health monitoring[119]. Courts may also learn from efforts to supervise remedies in the technology space, for example technical committees established in antitrust cases including the 2002 *Microsoft* antitrust consent decree[120,121] or as envisioned in the recent *Epic v. Google* case[122].

Because data collected via the external mechanism would be fully opt-in by user participants and explicitly collected with at-time-of-collection disclosures, researchers have discretion at how to implement and set standards. By contrast, full access to the private data of users of the company would be excessive outside of a proactive notification and acceptance process; however, researchers must have sufficient ability to audit results internal to the operation and confirm that audience definitions, specifications, and data collection match the research agenda approved by the court. That likely would require targeted review of user-level data; however, such data could be limited in distribution and would not need to leave company systems, assuming adequate access for the oversight team, with reporting and readouts (including public reports) of only aggregate data.

Critically, because many of the same questions will be asked by both the internal and external systems, the setup described here enables longitudinal assessments to be benchmarked against each other using those metrics. This is a critical advantage: it would allow for the normalization of metrics that appear within only one of the systems using the ones that do appear in both, ensuring that the internal workings of the company's systems are not at risk of biasing effects.

With implementation of the type described here, the research group would be positioned to deliver back to the court a periodic update on remediation, matching the normal engineering cycles of large technology companies. Given the complexity and multidimensional nature of benchmarking and oversight, the research group would also establish a proactive and regular process for stakeholder communication and consultation that aligns with timelines for updates to the court. Assuming monitoring and oversight occurs for a multi-year period, this operation would be established to work independently, with some ability to give quicker feedback to a judge in the event of meaningful implementation blockers, especially with the internal mechanism.

**Specific Judicial Mitigation**

Based on the specifics of a case and any already-established harms, it may be appropriate for a court to establish specific product requirements at the time of the initial settlement or ruling. We do not mean to limit the importance of this potential, but because this type of intervention depends greatly on the details of a case, we limit a deeper discussion here.

Depending on those details, candidate interventions may include mechanisms like requirements for stronger age verification, more restrictive default privacy settings, default limits on algorithmic optimization for usage, novel user control mechanisms, know your customer procedures for new account creation, monitoring practices for the most influential accounts, or procedures around countering trafficking and CSAM. Product changes stemming from proposed design codes[123], several US state laws[124], and the EU each offer potential approaches that may be appropriate[11]. Companies have made many changes that mirror these approaches, demonstrating both feasibility and perceived efficacy[93].

Of course, courts may also order remediation in the form of non-product requirements. These interventions may include educational or mental health support, including external funding, or any number of other specific actions to mitigate harms. We focus here on product interventions and especially the assessment of those interventions, but anticipate that the particulars of a case would very likely include a range of other interventions as well.

## METRICS

Both internal and external mechanisms require measurements to evaluate the harms of social media on users. As depicted in Figure 1, a range of potential methods can be drawn upon. A goal of this approach is to establish behavioral and directional metrics that are sensitive at the level of operational intervention while also connecting to the established harms affecting a population.

Internal and external mechanisms can utilize existing measurement tools to monitor the impacts of social media use. These tools, developed by social media companies and academic researchers, assess a broad spectrum of outcomes ranging from immediate outcomes such as user behavior, user experience and psychological responses, to consequences that develop and manifest over a longer term, including behavioral change and health outcomes (see details in Appendix 1). In general, social media companies focus mainly on short-term outcomes, such as encountering policy-violating content or having negative experiences on a platform (e.g. Meta's BEEF survey)[87], whereas independent researchers develop measurements to assess longer-term outcomes, such as mental health and wellbeing[125]. Therefore, in line with the existing measurements, we recommend adding metrics that measure intermediate indicators, including user experience, psychological/neurocognitive responses, and behavioral risk factors, through both internal and external mechanisms, while reserving the assessment of long-term health metrics for external evaluations.

To further guide the design of internal and external harm measurements, we offer a detailed social media harm taxonomy based on a review of existing legislation and litigation[93]. In the taxonomy (see Appendix 2), we identify four categories of social media harms—(1) unwanted or harmful content, (2) unwanted or harmful usage, (3) unwanted or harmful contact, (4) unwanted usage of user information and likeness—and three harm measurement methods—(1) user experience surveys, (2) content or behavior metrics in industry, and (3) empirical academic measures.

Almost all harms have both objective and subjective components to them, suggesting a need for both user experience and objective assessments of behavior and content. To monitor unwanted or harmful content and contact such as witnessing or being targeted by bullying, we recommend that both internal company data systems and external researchers conduct user experience surveys, with the internal systems also reporting objective metrics for comparison against survey results. This would allow for a comparison of how company definitions of harm map to user experiences. For surveillance of unwanted or harmful usage, internal company data systems ought to track and disclose objective user behaviors, such as screen time or sleep displacement, and provide external researchers with access to data samples for empirical analysis of these behaviors and their long-term impacts. This ought to be supplemented by user experience surveys that indicate when a user feels that their usage is or is not excessive and getting in the way of their goals, which is a hallmark of problematic usage. To oversee cases where user information might be misused, both the internal system and external researchers should utilize user experience surveys, since it may not be possible to understand welcome and unwelcome usage of one's information. In all cases, metrics collected should prioritize groups identified as vulnerable by prior research and compare the magnitude of harm in these specific groups to the holdout groups.

Utilizing a dual system—internal company systems and external assessment—to monitor various types of social media harms effectively addresses significant challenges within the current infrastructure. First, the slow development and manifestation of long-term outcomes present substantial difficulties in estimating complete causal effects of the harms. Moreover, the constantly-evolving design of social media platforms complicates causal inference in obtaining counterfactual scenarios[95,126]. Literature suggests that metrics derived from user experience surveys are predictive of long-term outcomes[127,128]. Therefore, conducting user experience surveys, such as the BEEF survey, within companies, not only meets the need to monitor intermediate harms of product designs but is also feasible within the existing company infrastructure.

Additionally, while user experience surveys, particularly the single-question self-reported measures (e.g., those used in the BEEF survey like "have you ever seen anyone try to bully others?") are generally less robust than those validated in academic research (e.g., the Cyber Violence Scale[129]), we recommend that external research employ harmonized user experience surveys. This approach ensures that employing the same metrics as those used in internal company surveys leads to comparable results, laying a solid foundation for monitoring the impacts of social media product designs. External researchers can administer these surveys to both sample groups shared by the social media company and other accessible users, thus enhancing the scope and diversity of monitoring, which, in turn, broadens harm evaluation.

**DISCUSSION**

This paper has sought to address some of the difficulties in causally measuring harms[95,126] by leveraging otherwise-difficult-to-implement structures that a court could require as part of either a mandated abatement or settlement processes. Once established, the core system implemented for one case might also be used for a range of other social media cases as well. Even and especially with data protection rules protecting internal user data along the lines used in the internal mechanism, a program such the one described here could establish teams and best practices that could make future monitoring easier and more comparable, similar to monitoring that was implemented for tobacco[60]. The external mechanism, especially, could even more easily be extended with additional scoping and resources, strengthening standards and enabling additional cross-platform comparisons.

Several ethics questions are raised in the context of a large-scale monitoring protocol. First, user privacy and participation in court-ordered experimentation is a persistent question. Significant parts of the protocol described here follow purely opt-in procedures (though still requiring robust disclosures and ex post data protection). For on-platform observation, especially within the internal mechanism, however, courts will need to be comfortable embracing rules that operate within the context of experiments that platforms already routinely perform, something quite different from the government standards associated with health trials or academic standards of institutional review boards (IRB). The existing operations of platforms may pose challenges for external academic researchers, whose university ethics standards may be stronger, or at least different, from those in industry. This area demands further exploration, including new academic approaches suitable for large-scale observational studies where individual opt-in may be impossible even while individual protections are critical. For now, we believe a principle of minimized individualized data access for researchers, limited to audit access to ensure compliance and data fidelity, is an appropriate approach and one grounded in some recent industry/academic collaborations[130]—though more work here is needed in preparation for the implementation of external collaboration for abatement as well as a range of other technical assessments of technology.

Lastly, ongoing legislative and regulatory updates may afford additional procedural strategies or implementation responses for abatement. For instance, the EU's forthcoming rules for enabling independent researcher access to platform data for regulatory supervision and the evaluation of systemic risks under the Digital Service Act's Article 40 may yield new research and data that could inform new approaches, as well as potential implementation partners, for abatement[131]. Similarly, should American state or federal law introduce further capacity for oversight, such rule shifts certainly could alter specific recommendations for abatement.

## CONCLUSION

In this article, we propose an implementable mechanism for abatement and monitoring in response to currently-pending or potential litigation related to social media companies. Should these cases reach settlement or a ruling in favor of the plaintiffs, a judge may use this approach as a base template to be updated with the specifics of any case. This proposal is situated to follow legal precedent in prior product liability cases, to incorporate standards from the public health literature, and to be implementable using the systems and practices that technology companies already employ. In particular, the mechanism utilizes a two-pronged structure: the internal mechanism affords access to experiments run internal to the platform, as well as associated metrics and decision-making processes, and to add experimental metrics associated with user harm; a second external mechanism would conduct user experience research and benchmark that can be compared over time against internal data as well as longer term well-being outcomes, and in comparison to other products. With court-appointed researchers managing this process, and with the ability to benchmark the two systems against one another, such a system would enable regular benchmarking and standards for improvement to ensure the mitigation of harm in response to the particular fact pattern established in a specific case.

## DISCLOSURES

Nathaniel Lubin, S. Bryn Austin, and Zachary J. Ward have been retained as experts to consult with an attorney general in social media litigation. Amanda Yarnell has received research grant support from Google.

## ACKNOWLEDGEMENTS

The authors would like to thank the feedback and suggestions from Nora Benavidez, Jacob Noti-Victor, Amanda Lenhart, Meetali Jain, Thomas Gilbert, Rob Leathern, and Brandon Silverman, among others.

**APPENDICES**

**Appendix 1: existing measurements of social media harms**

To monitor users' activities and experiences on social media effectively, both internal and external mechanisms require consistent and practical measurement tools. As illustrated in Figure 1, the potential impacts of social media on users encompass both short-term and long-term outcomes.

For instance, if an adolescent enabled personalized algorithms, they may receive a plethora of engaging content tailored to their interests. While initially receiving such content does not pose a mental health risk, it may extend usage time (user behavior) due to its compelling nature. Meanwhile, the sensational and provocative content recommended by algorithms may expose the user to extreme content or an overload of social posts from peers (user experience). Gradually, such exposure can elicit psychological responses, such as negative social comparisons, polarized views, or body image concerns (psychological responses). These psychological reactions may influence daily behaviors, such as participating in extreme events, engaging in body surveillance, or adopting late-night social media habits (behavioral risks). The accumulation of these factors can diminish overall well-being or contribute to the development of health issues (disease outcome or well-being). From this hypothetical example, it is clear that measuring the harms of social media must consider both the short-term outcomes, like changes in user behavior or experience, and those that develop and manifest over time, such as behavioral risks and health outcomes.

- User behavior: Although an external system can collect limited data on (self-reported) user behavior, including these indicators could be useful for benchmarking the internal and external systems. Importantly, there should also be a mechanism to link individual-level data collected from the external system to internal log/usage data from a platform, potentially using an "honest broker" system or other privacy-protecting approaches such as federated learning. This would improve the inferential capabilities of the external system to assess how platform usage patterns impact longer-term health outcomes.

- User experience: Similarly, asking questions about user experience (e.g. BEEF-style questions) in both the internal and external systems will allow for benchmarking across the systems and cross-platform comparisons from the external system.

- Psychological/neurocognitive responses: A range of indicators should be considered for responses that are presaged by user behavior and experience, such as emotional reactions (e.g. anxiety, fear of missing out—FOMO, colloquially), self-esteem and body image (e.g., body dissatisfaction), perceived social connectivity, motivation for social media use, and reports of addictive behaviors.

- <u>Behavioral risk factors:</u> A set of behavioral risk factors should also be assessed, such as indicators related to sleep (quality and duration), eating/dieting habits (including use of diet pills, etc.), self-harm, suicidal ideation, substance use (tobacco, drugs, alcohol), risky sexual behavior, and physical activity.

- <u>Disease screening questions:</u> A number of validated screening questions exist for mental health outcomes, including the Patient Health Questionnaire (PHQ)-2 and PHQ-9 for depression,[12] the SCOFF questionnaire for eating disorders,[13] and the Generalized Anxiety Disorder 7-item (GAD-7) for anxiety.[14]

- <u>Disease prevalence (diagnosed):</u> In addition to screening questions for (potentially undiagnosed) disease/health outcomes, a set of questions asking about diagnosed conditions would provide data on prevalence of harms that have been detected (and treated). Questions such as: "Has a doctor ever diagnosed you with depression?", and "Are you currently taking any medications for depression", or "Has a doctor, nurse, or other health professional ever told you that you have a depressive disorder including depression, major depression, dysthymia, or minor depression?"[15] provide a relatively simple approach to assessing diagnosed harms, and aligns with other disease estimation approaches such as the CDC's Behavioral Risk Factor Surveillance System.[16]

- <u>Subjective well-being:</u> Lastly, questions to elicit subjective well-being should be included to provide a more holistic view of health outcomes. Standard instruments such as the SF-6D and EQ-5D are routinely used,[17] and have the advantage of being mapped to utilities that can be used for economic evaluation analyses to estimate the cost-effectiveness of various strategies to improve social media-related health outcomes. Other instruments that may be more appropriate for children and adolescents, such as the PedsQL should also be considered.[18]

**Appendix 2: a taxonomy of social media harm measures**

We have identified four categories of social media harms and three types of measurements commonly used in both industry and academia to monitor these harms. For each harm category, we summarize many existing measurements and highlight those that are readily-applicable across various settings in red. Our aim is to provide professionals and researchers with examples of previously-used tools (Table A1 - Table A4) to monitor these harms and to encourage discussions about the framework of social media harms. While we strive to be comprehensive in the review, we acknowledge that due to the extensive literature in this field, there are likely many additional measurements not included here. We would welcome reader suggestions as to other measures to include and plan to publish periodic online updates of measures corresponding to these harms.

The four categories of social media harms include:

1. Unwanted or harmful content: exposure to potentially damaging content such as misinformation, violence, nudity, and extreme material (See: Table A1).
2. Unwanted or harmful usage: users developing negative or harmful psychological reactions to social media use, such as negative social comparison and fear of missing out; and users develop in response to social media use, such as loss of self-control and problematic use (See: Table A2).
3. Unwanted or harmful contact: harmful interactions from other users or bots, such as bullying or unwanted sexual advances (See: Table A3).
4. Unwanted usage of user information: threats to users' data and accounts on social media platforms, including privacy invasion, impersonation, and exposure to malware or adware (See: Table A4).

These measures have been measured in both industry and academia via three methods:

A. User experience surveys: Surveys inviting users to self-report their social media related experience, which can be measured both externally and internally. We recommend tracking these for every harm externally and independently from platforms, to be able to benchmark against internal metrics.
B. Content or behavior metrics in industry: Metrics developed by teams focused on trust, safety, and/or integrity at social media companies to monitor their progress against the prevalence of violating content and behavior. The content or behavior metrics can be reported by industry in relation to product changes, and benchmarked against user experience surveys.
C. Empirical academic research: Measures developed by academic researchers to investigate the level of social media harms among study participants, which can be benchmarked against user experience surveys and industry metrics.

**Table A1. Social media harm measures for <u>unwanted or harmful content</u>**

| Harm Description | By User Experience Surveys | By content or behavior metrics in industry | By empirical academic research* |
|---|---|---|---|
| Drug Sales | • "seen anyone trying to buy or sell" (BEEF*[19] Q5) | • Content based prevalence of violation of drug sales policy<br>• Reports of violations of drug sales policy | |
| Content that Encourages Self Harm | • "seen anyone harm themselves, or threaten to do so" (BEEF Q13) | • Content based prevalence of violation of self harm policy<br>• Reports for violations of self harm policy | |
| Misinformation | • "seen anything that was false or misleading" (BEEF Q1) | • Prevalence of Content Rated False by Fact Checkers<br>• User Reports of Misinformation (likely biased, so requires aggregation across diverse groups, such as is used by X's Community Notes feature) | |
| Witnessing Hate or Bullying | • "seen anyone discriminating against people" (BEEF Q3)<br>• "seen anyone do any of these things to someone else: insult / disrespect / threaten.." (BEEF Q4) | • Content Level Predictions of Toxicity (Perplexity API[20])<br>• Prevalence of Content that Violates Hate Speech or Bullying Policies | • Cyber-violence scale (Šincek 2021[21])<br>• Online racism scale (Keum 2021[22]) |
| Graphic/Violent Content | • "seen any violent, bloody, or disturbing images" (BEEF Q2) | • Image classification based predictions of prevalence.<br>• Prevalence based on violations of policies.<br>• Hiding or negative feedback on content. | |
| Unwanted Sexually Explicit Content | • "seen nudity or sexual images that you didn't want to see" (BEEF Q7) | • Image classification based predictions of prevalence.<br>• Prevalence based on violations of policies.<br>• Hiding or negative feedback on content. | |

* Note: We acknowledge our limitations in summarizing harm measurements from empirical research, as we have not conducted systematic reviews across all related fields. We welcome suggestions for improvement and will continue updating this table. The latest version is available here: https://docs.google.com/spreadsheets/d/1pqaBBgf6ZIdLyGHRuapg1WeCJ6NUw6KIIrvQmMpZeZ8/edit?usp=sharing.

**Table A2. Social media harm measures for <u>unwanted or harmful usage</u>**

| Harm Description | By User Experience Surveys | By content or behavior metrics in industry | By empirical academic research* |
|---|---|---|---|
| Excess Usage | "Excessive computer/internet use...have this issue?" (UMN, 2024[23]) | Time spent on social media including for heavy users (90th percentile) | |
| Negative Social Comparison | "felt worse about your life" (BEEF[24] Q8) | (None) | <ul><li>Negative social media comparison scale (Samra et al., 2022[25])</li><li>FOMO scale (Abel et al., 2016[26])</li><li>Appearance-related social media consciousness scale (Choukas-Bradley et al., 2020[27])</li><li>Social media rumination scale (Parris et al., 2022[28])</li><li>Digital stress scale (Hall et al., 2021[29])</li></ul> |
| Displacement of Sleep | "takes time away from other activities you care about" (Common Sense Media, 2024) | Time spent on social media at night including for heavy users (90th percentile) | |
| Control (Positive or Negative) | <ul><li>"ever felt a lack of control over what you see" (BEEF 21)</li><li>"can't control their use or end up using [xxx] for a longer period of time than they originally wanted to" (Common Sense Media, 2024[30])</li></ul> | Time spent on social media including for heavy users (90th percentile) | |
| Feeling Manipulated | "believe social media companies manipulate users to spend more time on their devices" (Common Sense | Time spent on social media including for heavy users (90th percentile) | Measures around harmful SMU psychological processes: <ul><li>Problematic social media use scales (Moretta et al., 2022[32])</li><li>Social media disorder scale (Van Den Eijnden et al., 2016[33])</li></ul> |

This pre-publication paper has been submitted for peer review and publication in a special issue of Annals of the New York Academy of Sciences on digital platforms and public health.

| | Media, 2018[31]) | | • Social media mindset and agency (Lee et al., 2023[34]) Measures of SMU related behavioral outcomes: • Sleep quality index (Buysse et al., 1991[35]) • Self harm screening inventory (Kim et al., 2022[36]) • Objectified Body Consciousness Scale (McKinley et al., 1996[37]) |

\* Note: We acknowledge our limitations in summarizing harm measurements from empirical research, as we have not conducted systematic reviews across all related fields. We welcome suggestions for improvement and will continue updating this table. The latest version is available here:

https://docs.google.com/spreadsheets/d/1pqaBBgf6ZIdLyGHRuapg1WeCJ6NUw6KIlrvQmMpZeZ8/edit?usp=sharing.

**Table A3. Social media harm measures for <u>unwanted or harmful contact</u>**

| Harm Description | By User Experience Surveys | By content or behavior metrics in industry | By empirical academic research* |
|---|---|---|---|
| Being Targeted for Bullying | "anyone done any of these things to you: insult / disrespect / threaten.." (BEEF[38] Q5) | <ul><li>Reports of Bullying</li><li>Content Level Predictions of Toxicity (Perplexity API[39])</li><li>Prevalence of Content that Violates Hate Speech or Bullying Policies</li></ul> | <ul><li>Online victimization scale (Tynes et al., 2010[40])</li></ul> |
| Fake Account Contact | "ever been contacted by an account that seemed fake" (BEEF Q14) | Prevalence of accounts that violate fake account/authenticity policies | |
| Unwanted Sexual Advances | "ever received unwanted sexual advances" (BEEF Q9) | <ul><li>Blocks of accounts</li><li>Reports of accounts for violating solicitation policies</li></ul> | <ul><li>Online sexual victimization (Gámez-Guadix et al., 2015[41])</li><li>Online sexual harassment scale (Buchanan et al., 2022[42])</li></ul> |

* Note: We acknowledge our limitations in summarizing harm measurements from empirical research, as we have not conducted systematic reviews across all related fields. We welcome suggestions for improvement and will continue updating this table. The latest version is available here:
https://docs.google.com/spreadsheets/d/1pqaBBgf6ZIdLyGHRuapg1WeCJ6NUw6KIIrvQmMpZeZ8/edit?usp=sharing.

**Table A4. Social media harm measures for <u>unwanted usage of user information</u>**

| Harm Description | By User Experience Surveys | By content or behavior metrics in industry | By empirical academic research* |
|---|---|---|---|
| Impersonation | "ever found out that a [service] account was pretending to be you" (BEEF[43] Q12) | (None) | |
| Data Privacy | "had concerns about how [service] might use data and information about you" (BEEF Q20) | (None) | |
| Misuse of your Image / Likeness | • "someone using my photos in an inappropriate way" (Savoia, et al., 2021[44])<br>• "ever send intimate or suggestive pictures/videos to someone online then later find out the person was not who they claimed to be/not who you thought they were" (Snapchat, 2023[45]) | (None) | Information privacy concerns inventories (Bartol et al., 2023[46]) |

* Note: We acknowledge our limitations in summarizing harm measurements from empirical research, as we have not conducted systematic reviews across all related fields. We welcome suggestions for improvement and will continue updating this table. The latest version is available here: https://docs.google.com/spreadsheets/d/1pqaBBgf6ZIdLyGHRuapg1WeCJ6NUw6KIIrvQmMpZeZ8/edit?usp=sharing.

Commission - Have Your Say.

https://ec.europa.eu/info/law/better-regulation/have-your-say/initiatives/13817-Delegated-Regulation-on-data-access-provided-for-in-the-Digital-Services-Act_en